\documentstyle[12pt]{article}

\newcommand{\be}{\begin{equation}}
\newcommand{\ee}{\end{equation}}
\newcommand{\bea}{\begin{eqnarray}}
\newcommand{\eea}{\end{eqnarray}}

\newcommand{\unitm}{{\bf 1}}

\begin{document}

\baselineskip 18pt

\begin{titlepage}
\begin{flushright}
CALT-68-2138 \\
September 1997
\end{flushright}

\vskip 1.2truecm

\begin{center}
{\Large {\bf Spin-Orbit Interaction 
   from Matrix Theory}}
\end{center}

\vskip 0.8cm

\begin{center}
 {\bf Per Kraus}$^1$
\vskip 0.3cm
{\it California Institute of Technology \\
     Pasadena CA 91125, USA  \\
     email: perkraus@theory.caltech.edu}
\end{center}

\vskip 2.2cm

\begin{center}
{\small {\bf Abstract: }}
\end{center}       
\noindent           
{\small 
We study the leading order spin dependence of graviton scattering in
eleven dimensions, and show that the results obtained from supergravity and
from Matrix Theory precisely agree.  }
\rm
\vskip 3.4cm

\small
\begin{flushleft}
$^1$ Work supported in part by DOE grant DE-FG03-92-ER40701 and by a DuBridge
fellowship.
\end{flushleft}
\normalsize 
\end{titlepage}

\newpage
\baselineskip 16pt

\section{Introduction}

There are by now a substantial number of checks of the correspondence
between Matrix Theory \cite{BFSS} and eleven dimensional supergravity.  
Impressive 
though they are, these checks have probed only a very limited part of the 
structure of supergravity.  Among the most striking of the checks are
the successful computations of the $v^{4}/r^{7}$ and $v^{6}/r^{14}$ terms
arising in the scattering of two gravitons\cite{BFSS,DKPS,BecBec,BBPT}.  
The results can be thought of
as probing the cubic and quartic interaction vertices of supergravity.  
However, these calculations effectively average over the polarizations of
the scattered gravitons, and so are sensitive only to the magnitude of the
vertices and not to their tensor structures.  In this work we will try to
see whether the tensor structure comes out correctly by studying the leading
order spin dependence of graviton scattering.

There have recently been some  string theory analyses of the spin dependence
of D0-brane scattering, or equivalently, graviton scattering in compactified
M theory.  \cite{Har} used a series of duality transformations to map
the problem to one involving fundamental strings, and \cite{MorScrSer}
approached the problem through the boundary state formalism. Here we
proceed somewhat differently, by finding the linearized metric of a spinning
D0-brane and studying the the action of a D0-brane probe moving in that
background.  Since we will be working with the linearized theory, we will
only pick up terms of first order in the spin, whereas  \cite{Har,MorScrSer} 
found the higher order contributions as well.  In the present approach, by
solving the full field equations it should be straightforward to recover the
extra terms, but we will not attempt that here.  

Our approach has the advantage that it can be extended to include 
contributions which are of higher orders in the gravitational coupling.
The problem with the other methods is that to compare with Matrix Theory
beyond the lowest order one must consider not standard IIA theory, but 
rather the theory resulting from compactifying M theory along a null direction
\cite{Nfin}.
In our framework this can easily be implemented by lifting the solution
to eleven dimensions and then recompactifying along a null direction.  It is
less clear how to proceed within the framework of \cite{Har,MorScrSer}.

On the Matrix Theory side, we will compute one loop contributions to the
effective action in the presence of both bosonic and fermionic background
fields.  The fermions encode the spin of the D0-brane in precisely the right
way to reproduce the supergravity result.  The term we compute is part of 
the supersymmetric completion of the bosonic $(F_{\mu\nu})^4$ terms which
arise at one loop.  It would be nice to demonstrate that supersymmetry 
 is sufficient to fix the coefficients of the fermionic terms.

Before proceeding to the calculations, let us mention that in principle an
efficient way of approaching the supergravity side of the problem would be
through use of the supersymmetric Born-Infeld action \cite{Sch,Ced,BerTow}.  
The coupling of 
a supersymmetric D0-brane to target space fields has been worked out in
\cite{Ced,BerTow}.  However, the results there are presented in 
terms of superfields,
whereas to apply them to the problem we are studying one really needs to work
out their component forms.

\section{Spin-orbit potential from supergravity}

We would like to compute, in ten dimensions, the linearized metric produced 
by a D0-brane of
mass $T_0$ and angular momentum $J^{ij}$.  It is easiest to begin by
considering the Einstein metric; formulas for the linearized metric are then
found in \cite{MyePer}.  Writing 
$$
g_{\mu\nu}^{E}=\eta_{\mu\nu}+h_{\mu\nu}^{E},
$$
with $\eta_{\mu\nu}={\rm diag}(-,+,+\cdots +)$, the needed formulas are
\begin{eqnarray}
h_{00}^{E}&=&\frac{16\pi G_{N}}{8\omega_{8}}\frac{T_{0}}{r^7} \nonumber \\
h_{ij}^{E}&=&\frac{16\pi G_{N}}{56\omega_{8}}\frac{T_{0}}{r^7} \delta_{ij} \\
h_{0i}^{E}&=&-\frac{8\pi G_{N}}{\omega_{8}}\frac{x^{k}J^{ki}}{r^9}. \nonumber
\end{eqnarray}
where $\omega_{8}$ is the area of the unit 8-sphere.  It is convenient to
introduce the quantity
$$
Q_{0}=\frac{16\pi G_{N}}{7\omega_{8}}T_{0}=\frac{15}{2T_{0}} .
$$
Note that we are working in string units ($2\pi \alpha'=1$).
Then,
\begin{eqnarray}
h_{00}^{E}&=&\frac{7}{8}\frac{Q_{0}}{r^7} \nonumber \\
h_{ij}^{E}&=&\frac{1}{8}\frac{Q_{0}}{r^7} \delta_{ij} \\
h_{0i}^{E}&=&-\frac{7}{2}\frac{Q_{0}}{T_{0}}\frac{x^{k}J^{ki}}{r^9}. \nonumber
\end{eqnarray}
Now we transform to the string metric using,
$$
g_{\mu\nu}^{S}=e^{\phi/2}g_{\mu\nu}^{E}.
$$
To linear order, the dilaton takes the same value as it does in the unspinning
case,
$$
e^{-\phi}=1-\frac{3}{4}\frac{Q_{0}}{r^{7}}.
$$
Thus we find,
\begin{eqnarray}
g_{00}^{S}&=&-\left(1-\frac{Q_{0}}{2 r^{7}}\right) \nonumber \\
g_{ij}^{S}&=&\left(1+\frac{Q_{0}}{2 r^{7}}\right)\delta_{ij} \\
g_{0i}^{S}&=&-\frac{7}{2}\frac{Q_{0}}{T_{0}}\frac{x^{k}J^{ki}}{r^9}. \nonumber 
\end{eqnarray}
$g_{00}^{S}$, $g_{ij}^{S}$ of course take their standard values, while 
 $g_{0i}^{S}$ gives the spin contribution.  

Actually, the metric that we want
is obtained from the one above by dropping the 1's in the parentheses of 
$g_{00}^{S}$, $g_{ij}^{S}$.  This is because we want a solution corresponding
to the theory obtained by compactifying eleven dimensional supergravity
along a null direction.  Such a solution can be generated by lifting the
metric (3) to eleven dimensions using the standard formulas for spacelike
compactification, and then returning to ten dimensions by compactifying a 
null direction.  The result is precisely to remove the 1's just mentioned.
In fact, this procedure is unecessary in the present context, as it only 
affects terms of higher power in velocity than the spin-orbit term, but
we mention it here for completeness.  From now on we drop the 1's and refer
to the resulting metric as simply $g_{\mu\nu}$.

Next, we consider the action of a D0-brane probe moving in this background:
\be
S_{0}=-T_{0}\int \! dt \ \left\{e^{-\phi}
\sqrt{-g_{\mu\nu}\dot{X^{\mu}}\dot{X^{\nu}}}-C_{\mu}\dot{X^{\mu}}\right\}.
\ee
To compute the potential we need to know the value of the RR gauge field
$C_{\mu}$.  We take
$$
C_{0}=-\frac{Q_{0}}{r^7} \ \ \ \ ; \ \ \ \ C_{i}=\frac{7}{4}\frac{Q_{0}}{T_0}
\frac{x^{k}J^{ki}}{r^9}.
$$
The form of $C_{0}$ is the conventional one.  The value for $C_{i}$ could be
arrived at by determining the magnetic moment of the D0-brane and using 
the standard formula for the resulting magnetic field.  Instead we have 
chosen the coefficient $7/4$ so as to cancel a term linear in velocity
coming from the expansion of $\sqrt{-g\dot{X}\dot{X}}$.  The linear term is
known to be absent (from the calculations of \cite{Har,MorScrSer}, for 
example). At any rate, this term is irrelevant as far as the coefficient of
the spin-orbit term is concerned.  

Now, inserting the fields into $S_{0}$ and expanding in powers of velocity
we find,
\begin{eqnarray}
S_{0}&=&-T_{0}\int \! dt \ \left\{1-\frac{1}{2}\vec{v}\cdot\vec{v}
+\frac{7}{8}\frac{Q_{0}}{T_0}
\frac{(x^{i}J^{ij}v^{j})(\vec{v}\cdot\vec{v})}{r^9}
-\frac{Q_{0}}{8}\frac{(\vec{v}\cdot\vec{v})^2}{r^7}+\cdots\right\} \nonumber \\
\mbox{}&=&\int \! dt \ \left\{-T_{0}+\frac{T_0}{2}\vec{v}\cdot\vec{v}
-\frac{105}{16 T_0}
\frac{(x^{i}J^{ij}v^{j})(\vec{v}\cdot\vec{v})}{r^9}
+\frac{15}{16}\frac{(\vec{v}\cdot\vec{v})^2}{r^7}+\cdots\right\}.
\end{eqnarray}
The third term gives the spin-orbit term which we would like to reproduce
from Matrix Theory.

\section{Spin-orbit potential from Matrix Theory}

On the Matrix Theory side the calculation proceeds by evaluating the quantum
mechanical effective action of a system of two D0-branes.  Before doing any
explicit computation, dimensional analysis and the systematics of the loop
expansion allow one to write the general form of the effective action
\cite{BBPT,BerMin},
\be
S_{l}\sim\int\! dt \, r^{4-3l}f\left(\frac{v}{r^2},\frac{\psi}{r^{3/2}}\right).
\ee
Here $S_l$ denotes the $l$'th loop contribution.  Note that at the one loop
level the action includes the term $v^{3}\psi^{2}/r^{8}$, and that this term
has the correct $v$ and $r$ dependence to match onto the spin-orbit term
of (5).  We would like to check whether the numerical coefficient and the
tensor structure similarly agree.  

We will be following the conventions of \cite{BecBec}, and some details of
the calculation which we omit can be found there.  
The Matrix theory action, including gauge fixing and ghost terms, is
\be
S=\int \! dt \ {\rm Tr}\left\{\frac{T_{0}}{2}F_{\mu\nu}F^{\mu\nu}-
i\bar{\psi}\!\not\!\!{D}\psi+T_{0}(\bar{D}^{\mu}A_{\mu})^2\right\}
+S_{\rm ghost},
\ee
where $\mu, \nu = 0\ldots 9$, $A_{\mu}=(A,X_{i})$, and
\begin{eqnarray}
\bar{D}^{\mu}A_{\mu}&=&-\partial_{t}A+\left[B^{i},X_{i}\right] \nonumber \\
F_{0i}&=&\partial_{t}X_{i}+\left[A,X_{i}\right] \nonumber \\
F_{ij}&=&\left[X_{i},X_{j}\right] \\
D_{t}\psi&=&\partial_{t}\psi+\left[A,\psi\right] \nonumber \\
D_{i}\psi&=&\left[X_{i},\psi\right] \nonumber 
\end{eqnarray}
$B^{i}$ is the bosonic background field. The fluctuations about the background
will be denoted by $Y^{i}$,
$$
X^{i}=B^{i}+\frac{Y^{i}}{\sqrt{T_0}}.
$$
We will be studying a system of 
two D0-branes, so all fields take values in the Lie algebra of U(2).  
  In terms of the U(2) generators we write,
\begin{eqnarray}
A&=&\frac{i}{2}\left(A_{0}\,\unitm +A_{a}\sigma^{a}\right) \nonumber \\
X^{i}&=&\frac{i}{2}\left(X^{i}_{0}\,\unitm +X^{i}_{a}\sigma^{a}\right)  \\
\psi&=&\frac{i}{2}\left(\psi_{0}\,\unitm +\psi_{a}\sigma^{a}\right) \nonumber 
\end{eqnarray}
For the background fields, we will give nonzero values to
$$
B^{1}_{3}=vt, \ \ \ B^{2}_{3}=b \ \ \ {\rm and} \ \psi_{3}.
$$
The values given to $B^{i}_{a}$ correspond to two D0-branes moving with 
relative velocity $v$ along the $x^{1}$ direction, and separated by distance
$b$ along the $x^{2}$ direction.  

The fermionic background $\psi_{3}$ gives fermionic expectation value 
$\pm \psi_{3}/2$ to each of the two D0-branes.  However, one is free to shift
the background by an amount proportional to $\unitm$ since the U(1) part of
the action decouples from the SU(2) part.  Thus the setup equally well
applies to the case where the two D0-branes have fermionic expectation
values $\psi_{3}$ and $0$.  The latter picture corresponds to the supergravity
configuration we are trying to describe.

Now it is straightforward but tedious to expand out the action in terms of the
field components defined above.  We work in Euclidean space
$t\rightarrow i\tau, A\rightarrow -iA$.  The action takes the schematic 
form
\be
S\sim (A)^2 +(Y)^2 +(\psi)^2+\dot{B}AY+\psi_{3}Y\psi+\psi_{3}A\psi+\cdots,
\ee
where $\cdots$ indicates terms cubic and quartic in fluctuations which 
won't contribute to our analysis.  Our strategy will be to treat the first
four terms exactly and the last two terms perturbatively.  That is, in terms
of Feynman diagrams, the first four terms supply the propagators and the
last two will supply the vertices.  Let us first study the mass spectrum by
considering the propagator terms.  One finds the following mass eigenstates:

\begin{eqnarray}
Y^{n}_{\pm}&=&\frac{Y^{n}_{1}\pm iY^{n}_{2}}{\sqrt{2}} \ \ \ \ (n=2\cdots 9)
  \ \ \ \ \ \ \ m^{2}=r^2 \nonumber \\
S_{\pm}&=&\frac{Y^{1}_{1}\pm iY^{1}_{2}\mp iA_{1}+ A_{2}}{\sqrt{2}} \ \ 
  \ \ \ \ \ \ \ \ \ \      m^{2}=r^2+2v \\
T_{\pm}&=&\frac{Y^{1}_{1}\pm iY^{1}_{2}\mp iA_{1}- A_{2}}{\sqrt{2}}  \ \ 
 \ \ \ \ \ \ \ \ \ \ m^{2}=r^2-2v \nonumber \\
\psi_{\pm}&=&\frac{\psi_{1}\pm i\psi_{2}}{\sqrt{2}}  \ \  
 \ \ \ \ \ \ \ \ \ \ \ \ \ \ \ \  \ \ \ \ \ \ \ \ \ \ \ 
m=v\tau\gamma_{1}+b\gamma_{2} \nonumber
\end{eqnarray}
where $r^2=b^2+(v\tau)^2$.  Here, by ``mass eigenstate'' we mean that
the action takes the form
$$
i\int \! d\tau \ 
\frac{1}{2}\phi_{+}(\partial_{\tau}^{\,2}-m_{\phi}^{\,2})\phi_{-} 
 \  \ \ \ {\rm and} \ \ \ \
i\int \! d\tau \  \psi_{+}^{T}(\partial_{\tau}-m_{\psi})\psi_{-}
$$
for the case of bosons and fermions respectively.  In addition to the massive
fields just described, there are massless fields which play no role in the
following discussion.  

Given these quadratic actions, we can work out propagators.  For the bosonic
fields we define $\Delta_{B}(\tau_{1},\tau_{2}\mid m^2)$ as the solution to
\be
(-\partial_{\tau_{1}}^{\,2}+m^{2})\Delta_{B}(\tau_{1},\tau_{2}\mid m^2)
=\delta(\tau_1-\tau_2).
\ee
Note that $m$ is allowed to be time dependent.  Then we find,
\begin{eqnarray}
\langle Y_{-}^{n}(\tau_{1})Y_{+}^{n'}(\tau_{2})\rangle &=&
\Delta_{B}(\tau_{1},\tau_{2}\mid r^2)\ \delta^{nn'}  \nonumber \\
\langle S_{-}(\tau_{1})S_{+}(\tau_{2})\rangle &=&
\Delta_{B}(\tau_{1},\tau_{2}\mid r^2+2v)   \\
\langle T_{-}(\tau_{1})T_{+}(\tau_{2})\rangle &=&
\Delta_{B}(\tau_{1},\tau_{2}\mid r^2-2v). \nonumber
\end{eqnarray}
We similarly define $\Delta_{F}$ by
\be
(-\partial_{\tau_{1}}+m)
\Delta_{F}(\tau_{1},\tau_{2}\mid m)
=\delta(\tau_1-\tau_2).
\ee
Then
\be
\langle \psi_{+}(\tau_{1})\psi_{-}(\tau_{2})\rangle =
\Delta_{F}(\tau_{1},\tau_{2}\mid  v\tau_{1}\gamma_{1}+b\gamma_{2}) .
\ee
In fact,  $\Delta_{F}$ can be related to $\Delta_{B}$ by
\be
\Delta_{F}(\tau_{1},\tau_{2}\mid  v\tau_{1}\gamma_{1}+b\gamma_{2})
=(\partial_{\tau_{1}}+v\tau_{1}\gamma_{1}+b\gamma_{2})
\Delta_{B}(\tau_{1},\tau_{2}\mid r^2-v\gamma_{1}).
\ee
It will turn out that we won't need the full structure of $\Delta_{F}$, but
only part of it.  The formulas we will need are
\begin{eqnarray}
\psi_{3}^{T}P_{+}\Delta_{F}(v\tau_{1}\gamma_{1}+b\gamma_{2}) P_{-}\psi_{3}
&=&\frac{b}{2} \psi_{3}^{T}\gamma_{1}\gamma_{2}\psi_{3} \ 
\Delta_{B}(r^2+v) \nonumber \\
\psi_{3}^{T}P_{-}\Delta_{F}(v\tau_{1}\gamma_{1}+b\gamma_{2}) P_{+}\psi_{3}
&=&-\frac{b}{2} \psi_{3}^{T}\gamma_{1}\gamma_{2}\psi_{3} \ 
\Delta_{B}(r^2-v) \\
\sum_{n=2}^{9} 
\psi_{3}^{T}\gamma_{n}\Delta_{F}(v\tau_{1}\gamma_{1}+b\gamma_{2})\gamma_{n}
\psi_{3} &=&\frac{3b}{2} \psi_{3}^{T}\gamma_{1}\gamma_{2}\psi_{3} \ 
[\Delta_{B}(r^2+v)-\Delta_{B}(r^2-v)], \nonumber
\end{eqnarray}
where $P_{\pm}=(1\pm\gamma_{1})/2$, and we have suppressed the $\tau$ 
dependence.  These relations  are easily derived upon recalling 
$\psi_{3}^{T}\psi_{3}=\psi_{3}^{T}\gamma_{i}\psi_{3}=0$, which follows from
the grassmann property of $\psi_{3}$ and the symmetry of $\gamma_{i}$.  

Now we can work out the fermionic dependence of the one loop effective action.
For this, we need to first find the $\psi_{3}Y\psi$ and $\psi_{3}A\psi$
terms in $S_{\rm fermi} = -i\int \!  dt \, {\rm Tr}\,\bar{\psi}\!\not\!\!D\psi$.
We find,
\begin{eqnarray}
S_{\rm fermi}&=&-\frac{i}{\sqrt{T_{0}}}\int \! d\tau \ \left\{
Y^{n}_{-}\psi_{3}^{T}\gamma_{n}\psi_{+}+Y^{n}_{+}\psi_{-}^{T}\gamma_{n}\psi_{3}
 \right.
+\sqrt{2}S_{-}\psi_{3}^{T}P_{+}\psi_{+}-\sqrt{2}S_{+}\psi_{-}^{T}P_{-}\psi_{3}
\nonumber \\
 & & \left.
-\sqrt{2}T_{-}\psi_{3}^{T}P_{-}\psi_{+}+\sqrt{2}T_{+}\psi_{-}^{T}P_{+}\psi_{3}
\right\}.
\end{eqnarray}
The spin-orbit interaction is found by expanding $e^{iS_{\rm fermi}}$ to
quadratic order in $\psi_{3}$ and taking the vacuum expectation value.  This
gives
\begin{eqnarray}
S_{\rm so} = -\frac{1}{T_{0}}\int \!  d\tau_{1} \, d\tau_{2} \ \! \left\{ \right. 
\!\!\!\!\!\!\!\!\!  & & 
\langle Y^{n}_{-}(\tau_{1})Y^{n'}_{+}(\tau_{2})\rangle \,
\psi_{3}^{T}\gamma_{n}\langle\psi_{+}(\tau_{1})\psi_{-}(\tau_{2})\rangle
\gamma_{n'}\psi_{3}  \nonumber \\
& &
-2\langle S_{-}(\tau_{1})S_{+}(\tau_{2})\rangle \,
\psi_{3}^{T}P_{+}\langle\psi_{+}(\tau_{1})\psi_{-}(\tau_{2})\rangle
P_{-}\psi_{3}  \nonumber  \\
& &-2\langle  T_{-}(\tau_{1})T_{+}(\tau_{2})\rangle \,
\psi_{3}^{T}P_{-}\langle\psi_{+}(\tau_{1})\psi_{-}(\tau_{2})\rangle
P_{+}\psi_{3} 
\left. \right\}
\end{eqnarray}
Using our previous results for the propagators we obtain
\begin{eqnarray}
S_{\rm so} &=&-\frac{b}{T_{0}}\,\psi_{3}^{T}\gamma_{1}\gamma_{2}\psi_{3} 
\,\int \! d\tau_{1}\, d\tau_{2}  \left\{ \right.
 3\Delta_{B}(\tau_{1},\tau_{2}\mid r^2)\left[
\Delta_{B}(\tau_{1},\tau_{2} \mid r^2 + v) 
-\Delta_{B}(\tau_{1},\tau_{2} \mid r^2 - v)\right] \nonumber \\
&& \ \ -\Delta_{B}(\tau_{1},\tau_{2} \mid r^2 + 2 v) 
\Delta_{B}(\tau_{1},\tau_{2} \mid r^2 +v) 
+\Delta_{B}(\tau_{1},\tau_{2} \mid r^2 - 2 v) 
\Delta_{B}(\tau_{1},\tau_{2} \mid r^2 - v) \left. \right\} \nonumber
\end{eqnarray}
It is evident that the terms linear and quadratic in velocity will cancel out 
in the above
expression.  To evaluate the $v^3$ term we need to expand out the 
propagators and compute the integrals.  After doing the $\tau_{2}$ integral
the result will take the form
$$
b v^{3}\psi_{3}^{T}\gamma_{1}\gamma_{2}\psi_{3}\int \! d\tau_{1} \,
\frac{1}{(b^2+v^2\tau_{1}^{\,2})^{9/2}}.
$$
Given this fact, it is easier to proceed by evaluating the $\tau_{2}$
integral with $\tau_{1}=0$ and then restoring the $\tau_{1}$ dependence
afterwards.
The expansion of the bosonic propagator is
\begin{eqnarray}
\Delta_{B}(0,\tau_{2} && \!\!\!\!\!\!\! 
\mid b^2 +\alpha v)=\frac{e^{-b|\tau_{2}}|}{b}
\left\{\frac{1}{2}-\frac{\alpha v}{4b^2}[1+b|\tau_{2}|] \right. \nonumber  \\
&& \! \!\!\!\!\!\!\!\!\!\!\!\!\!\!\!\!
+\frac{v^2}{48b^4}\left[(9\alpha^2-6)(1+b|\tau_{2}|)+(3\alpha^2-6)b^{2}
|\tau_{2}|^2-4b^{3}|\tau_{2}|^{3}\right] \nonumber \\
&& \left.  \! \!\!\!\!\!\!\!\!\!\!\!\!\!\!\!\!
-\frac{\alpha v^3}{96b^6}\left[(15\alpha^2-30)(1+b|\tau_{2}|)
+(6\alpha^2-24)b^2|\tau_{2}|^2+(\alpha^2-14)b^{3}|\tau_{2}|^{3}
-4b^{4}|\tau_{2}|^{4}\right]\right\} + \cdots \nonumber
\end{eqnarray}
Plugging this expansion into $S_{\rm so}$ and doing the $\tau_{2}$ integral
gives
\be
S_{\rm so}= -\frac{105}{32T_{0}}\int \! d\tau_{1} \, 
\frac{b v^{3}\psi_{3}^{T}\gamma_{1}\gamma_{2}\psi_{3}}{r^{9}}.
\ee

Now we can compare with the result from supergravity.  Transforming back to
Minkowski space, we find that the spin-orbit terms from 
(5) and (20) agree provided
\be
\frac{(x^{i} J^{ij} v^{j})(\vec{v} \cdot \vec{v})}{r^{9}}=
-\frac{i}{2}\frac{b v^{3}\psi_{3}^{T}\gamma_{1}\gamma_{2}\psi_{3}}{r^{9}}.
\ee
Does this equivalence make sense?  To see that it does we need to recall the
expression for the angular momentum operator of Matrix Theory.  Starting
from the action (7), the operator which generates rotations in the 
transverse space is the sum of a bosonic piece and a fermionic piece. If we
work in the rest frame of the source D0-brane - the one carrying the 
fermionic expectation value - then the bosonic contribution to the angular 
momentum of the source vanishes.  The 
fermionic piece is the standard expression for the angular momentum of a 
spinor field,
$$
J^{ij}=\frac{i}{2}\bar{\psi}\gamma_{i}\gamma_{j}\psi.
$$
Recalling that $\psi$ is Majorana, and that the relative velocity and 
separation
of the D0-branes are along the $x^1$ and $x^2$ axes respectively, we find that
(21) is satisfied.  Thus we have verified that supergravity and Matrix Theory
agree as to the leading spin dependence of the scattering amplitude.

\bigskip

{\Large {\bf Acknowledgements}}

I am grateful to M. Becker, E. Keski-Vakkuri, and J. Schwarz for helpful
discussions.  I would also like to thank M. Serone for helping me to correct
an error in an earlier version of the manuscript.

\end{document}